\documentclass[twocolumn,english,aps,prl]{revtex4-1}
\usepackage{color}
\usepackage{amsmath}
\usepackage{graphicx}
\usepackage{amssymb}

\makeatletter
\@ifundefined{textcolor}{}
{%
 \definecolor{BLACK}{gray}{0}
 \definecolor{WHITE}{gray}{1}
 \definecolor{RED}{rgb}{1,0,0}
 \definecolor{GREEN}{rgb}{0,1,0}
 \definecolor{BLUE}{rgb}{0,0,1}
 \definecolor{CYAN}{cmyk}{1,0,0,0}
 \definecolor{MAGENTA}{cmyk}{0,1,0,0}
 \definecolor{YELLOW}{cmyk}{0,0,1,0}
 }

\@ifundefined{definecolor}
 {\usepackage{color}}{}
\@ifundefined{definecolor}{\@ifundefined{definecolor}
 {\usepackage{color}}{}
}{}

\newcommand{\sgn}{\operatorname{sgn}}

\newcommand{\ba}{\begin{eqnarray*}}
\newcommand{\ea}{\end{eqnarray*}}
\newcommand{\baa}{\begin{eqnarray}}
\newcommand{\eaa}{\end{eqnarray}}
\newcommand{\bea}{\begin{eqnarray}}
\newcommand{\eea}{\end{eqnarray}}
\newcommand{\be}{\begin{equation}}
\newcommand{\ee}{\end{equation}}

\newcommand{\sm}{SmB$_6$}
\newcommand{\pu}{PuB$_6$}

\newcommand{\C}{\mathcal{C}}

\newcommand{\bk}{\mathbf{k}}

\DeclareMathOperator{\ii}{i}

\makeatother

\usepackage{babel}

\begin{document}

\title{
Distinct topological crystalline phases in models for the strongly correlated topological insulator {\sm}
}

\author{Pier Paolo Baruselli}
\author{Matthias Vojta}
\affiliation{Institut f\"ur Theoretische Physik, Technische Universit\"at Dresden, 01062 Dresden, Germany}


\begin{abstract}
{\sm} was recently proposed to be both a strong topological insulator and a topological crystalline insulator.
For this and related cubic topological Kondo insulators,
we prove the existence of four different topological phases, distinguished by the sign of mirror Chern numbers. We characterize these phases in terms of simple observables, and we provide concrete tight-binding models for each phase.
Based on theoretical and experimental results for {\sm} we conclude that it realizes the phase with $\C^+_{k_z=0}=+2$, $\C^+_{k_z=\pi}=+1$, $\C^+_{k_x=k_y}=-1$,
and we propose a corresponding minimal model.
\end{abstract}

\date{May 13, 2015}


\maketitle


Topological insulators (TIs) with strong electronic correlations are considered to be of crucial importance in the exciting field of topological phases: They may provide TI states which are truly bulk-insulating -- a property missing from many Bi-based TIs -- and they may host novel and yet unexplored interaction-driven phenomena.

In this context, the material {\sm} has attracted enormous attention: it has been proposed \cite{takimoto,lu_smb6_gutz, tki_cubic} to \mbox{realize} a three-dimen\-sional (3D) topological Kondo insulator (TKI). This is a system where $f$-electron local moments form at intermediate temperatures $T$ and are subsequently screened at low $T$, such that a topologically non-trivial band\-structure emerges from Kondo screening \cite{tki1}.


While the results of numerous experiments on {\sm}, such as transport  \cite{wolgast_smb6,fisk_smb6_topss, smb6_junction_prx} and quantum oscillation measurements \cite{lixiang_smb6} as well as angle-resolved photoemission spectroscopy (ARPES) \cite{neupane_smb6,mesot_smb6, smb6_arpes_feng,smb6_arpes_reinert,smb6_past_allen} and spin-resolved ARPES (SP-ARPES) \cite{smb6_arpes_mesot_spin},
appear consistent with the presence of Dirac-like surface states expected in a TKI, doubts have been raised about the proper interpretation of data \cite{sawatzky_smb6,smb6_prx_arpes,smb6_trivial}.
Clearly, both experimental and theoretical progress is required for a thorough comprehension of this material.

%

A recent theoretical insight \cite{smb6_tci} is that {\sm} is also a topological crystalline insulator (TCI) \cite{fu_tci}, having three non-zero mirror Chern numbers. While Ref.~\onlinecite{smb6_tci} concluded that the latter are, modulo 2 or 4, independent of model details, we show in this Letter that the individual values of these Chern numbers do depend on model details. As a result, we obtain four distinct TCI phases which in particular differ in their surface-state spin structure.
Using results from density-functional theory (DFT) and SP-ARPES we are able to single out one of the four phases as being relevant for {\sm}. We also show that some of the previously employed models \cite{tki_cubic,sigrist_tki_prb} do not belong to this phase, and we propose a new minimal model for the correct TCI phase.
Our results highlight the non-trivial role played by mirror Chern numbers in describing topological band structures of actual materials.

\begin{figure}[b]
\includegraphics[width=0.48\textwidth]{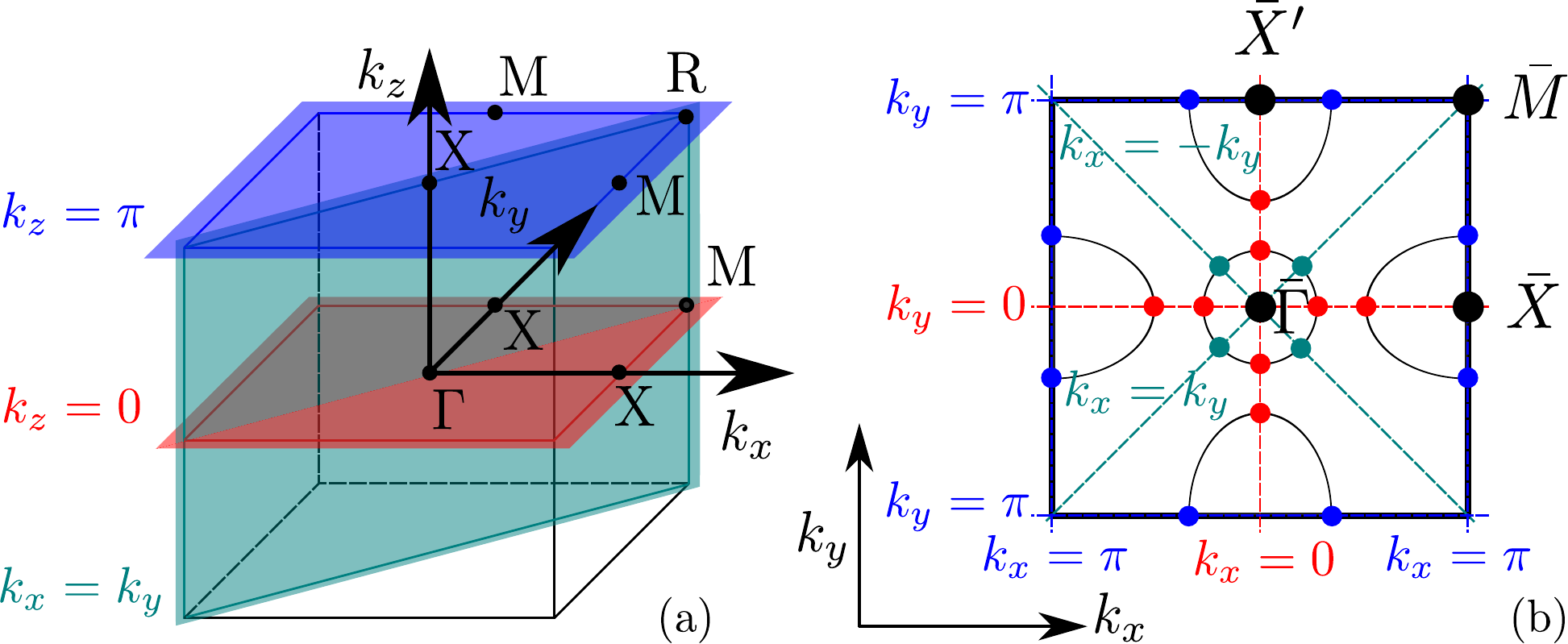}
\caption{(a) 3D BZ and its mirror planes ${k_z=0}$, ${k_z=\pi}$, ${k_x=k_y}$;
high-symmetry points are $\Gamma\!=\!(0,0,0)$, $X\!=\!(\pi,0,0), (0,\pi,0), (0,0,\pi)$, $M\!=\!(0,\pi,\pi),(\pi,0,\pi),(\pi,\pi,0)$, $R=(\pi,\pi,\pi)$.
(b) Corresponding 2D BZ for a (001) surface with
its high-symmetry points $\bar\Gamma\!=\!(0,0)$, $\bar{X}\!=\!(\pi,0)$, $\bar{X}'\!=\!(0,\pi)$ and $\bar{M}\!=\!(\pi,\pi)$ and its mirror planes ${k_{x,y}=0}$, ${k_{x,y}=\pi}$, ${k_x=\pm k_y}$,
which are pairwise equivalent by $C_{4v}$ symmetry. (Planes at $k_{x,y}=\pi$ and $k_{x,y}=-\pi$ are identical and share a single label.)
Also shown is the 2D Fermi surface of \sm, consisting of three pockets centered at $\bar\Gamma$, $\bar X$, and $\bar X'$. Dots mark its intersections with mirror planes.
}
\label{fig_3dbz}
\end{figure}

{\em Lattice and symmetries.}
{\sm} crystallizes in a simple-cubic ($sc$) structure, with its first Brillouin zone (BZ) shown in Fig.~\ref{fig_3dbz}(a). In the following we use the lattice spacing $a\!=\!4.13$\,{\AA} as unit length.
We may introduce mirror operators $M_l\equiv P C_2(l)$,
where $P$ is the inversion and $C_2(l)$ is a rotation by $\pi$ about the axis $l$ perpendicular to the mirror plane. For spin-$1/2$ particles $M_l^2=-1$, such that $M_l$ operators have eigenvalues $\pm \ii$.
For cubic symmetry there are three independent momentum-space planes which are invariant under the relevant mirror operators: the planes ${k_z=0}$ (equivalent to ${k_{x,y}=0}$) and ${k_z=\pi}$ (equivalent to ${k_{x,y}=\pi}$) are invariant under $M_z$ while the plane ${k_x=k_y}$ (equivalent to ${k_x=-k_y}$, ${k_y=\pm k_z}$, ${k_z=\pm k_x}$) is invariant under $M_{x-y}$; see Fig.~\ref{fig_3dbz}(a).



\begin{table*}
\centering
$$
\begin{array}{|l|c|c|c|c|c|c|c||c|c|c|c|c|c|}\hline
\mbox{Model} &\mbox{Abbr.} &N_{bands} &\mbox{Basis}&\multicolumn{4}{|c||}{\mbox{Parameters}} & \C^+_{k_z=0}& \C^+_{k_z=\pi} & \C^+_{k_x=k_y} & v&w&\mbox{Fig. \ref{fig_chern}}\\
\hline
\mbox{$\Gamma_8$ 1NN hyb.} 			&\Gamma_8^{1x}& 8  & E_g, \Gamma_8&\epsilon^f-\epsilon^d& \eta_z^{d1},\eta_z^{d2}&\eta_z^{f1},\eta_z^{f2}&\eta_x^{v1}& -2 & +1 &+1&-1&-1& (a)\\
\mbox{$\Gamma_7$ 2NN hyb.} 			&\Gamma_7^{2z}& 6  & E_g, \Gamma_7&\epsilon^f-\epsilon^d& \eta_z^{d1},\eta_z^{d2}&\eta_7^{f2},\eta_7^{f3}&\eta_7^{v2}& -2 & +1 &+1&-1&-1& (a)^*\\\hline
\mbox{$\Gamma_8$ 2NN hyb.} 			&\Gamma_8^{2z}& 8  & E_g, \Gamma_8&\epsilon^f-\epsilon^d& \eta_z^{d1},\eta_z^{d2}&\eta_z^{f1},\eta_z^{f2}&\eta_z^{v2} & +2 & +1 &-1&-1&+1& (b)\\
\mbox{$\Gamma_7$ 1NN hyb.}			&\Gamma_7^{1x}& 6  & E_g, \Gamma_7&\epsilon^f-\epsilon^d& \eta_z^{d1},\eta_z^{d2}&\eta_7^{f2},\eta_7^{f3}&\eta_7^{v1}& +2 & +1 &-1&-1&+1& (b)^*\\
\mbox{Full tight-binding \cite{pub6,prbr_io_smb6}} 	&\mbox{FTB}& 10 & E_g, \Gamma_8, \Gamma_7& \multicolumn{4}{|c||}{\mbox{many}}& +2 & +1 &-1&-1&+1& (b)\\ \hline
\mbox{Alexandrov \cite{tki_cubic} 1NN hyb.}	&A^{1z}& 8  & E_g, \Gamma_8&\epsilon^f-\epsilon^d& \eta_z^{d1},\eta_x^{d1}&\eta_z^{f1},\eta_x^{f1}&\eta_z^{v1} & -2 & +1 &-1&+1&-1& (c)\\
\mbox{Legner et al. \cite{sigrist_tki_prb} }		&sf& 4  & s, \Gamma_8^{(2)}&\epsilon^f-\epsilon^s&\eta^{s1},\eta^{s2}&\eta^{f1},\eta^{f2}&\eta^{v1}  & -2 & +1 &-1&+1&-1& (c)\\ \hline
\mbox{Alexandrov \cite{tki_cubic} 2NN hyb. }	&A^{2x}& 8  & E_g, \Gamma_8&\epsilon^f-\epsilon^d& \eta_z^{d1},\eta_x^{d1}&\eta_z^{f1},\eta_x^{f1}&\eta_x^{v2} & +2 & +1 &+1&+1&+1& (d)\\\hline
\end{array}$$
\vspace*{-15pt}
\caption{Mirror Chern numbers  $\C^+_{k_z=0}$, $\C^+_{k_z=\pi}$, $\C^+_{k_x=k_y}$ in different tight-binding models for {\sm}, with 1NN (2NN) referring to first (second) neighbor hybridization.
Also quoted are the number of bands $N_{bands}$, the orbital basis, and the non-zero terms in the Hamiltonian required by each model (divided into on-site energies $\epsilon$, $d$ hopping $\eta^d$, $f$ hopping $\eta^f$, hybridization $\eta^v$) \cite{model_parameters}.
The four phases, distinguished by
$v\equiv\sgn( \C^+_{k_z=0}\C^+_{k_x=k_y})$ and $w\equiv\sgn(\C^+_{k_z=0}\C^+_{k_z=\pi})$,
are illustrated in Fig.~\ref{fig_chern} (where for $\Gamma_7$ models$^*$ the spin directions needs to be reversed).
}\label{table_chern}
\end{table*}


{\em Topological invariants.}
For each momentum-space mirror plane, one can define a mirror Chern number \cite{smb6_tci}:
%
%
\be\label{chern_number}
\C^\pm_{\overline{BZ}}=\frac{\ii}{2\pi}\sum_{a,b=1}^2\epsilon_{ab}\sum_{n=1}^N\int_{\overline{BZ}}d^2\bk \langle \partial_a u_n^\pm(\bk)| \partial_b u_n^\pm(\bk)\rangle,
\ee
with $M|u_n^\pm(\bk)\rangle=\pm \ii|u_n^\pm(\bk)\rangle$ and $\bk$ lying in the plane $\overline{BZ}$  which is invariant under the symmetry operator $M$
($M$=$M_z$ when $\overline{BZ}$ is $k_z=0$ or $k_z=\pi$, $M=M_{x-y}$ when $\overline{BZ}$ is $k_x=k_y$),
and we sum over all $N$ occupied bands.
We note that $\C^+_{\overline{BZ}}+\C^-_{\overline{BZ}}=0$ and, by cubic symmetry, $\C^+_{k_z=0}=\C^+_{k_x=0}=\C^+_{k_y=0}$ etc.
A given insulating bandstructure is thus characterized by a triplet of numbers ($\C^+_{k_z=0}$, $\C^+_{k_z=\pi}$, $\C^+_{k_x=k_y}$).

We recall that, according to bandstructure calculations \cite{takimoto,lu_smb6_gutz}, {\sm} is a strong TI: Band inversion between $d$ and $f$ states is achieved at the three $X$ points of the 3D BZ, leading to the parity invariants \cite{kane_bisb} $\delta(\Gamma)=+1$, $\delta(X)=-1$, $\delta(M)=+1$, $\delta(R)=+1$
and to the $\mathbb{Z}_2$ indices $(\nu_0,\nu_1\nu_2\nu_3)=(1,111)$.
Ref.~\onlinecite{smb6_tci} proved, using the properties of the wavefunctions at high-symmetry points, that the mirror Chern numbers of {\sm} obey $\C^+_{k_z=0}=2 \mod 4$, $\C^+_{k_z=\pi}=1 \mod 4$, $\C^+_{k_x=k_y}=1 \mod 2$ {\em independent} of microscopic details (like the $\mathbb{Z}_2$ indices).
Below we show this does {\em not} apply to the individual values of these Chern numbers which instead are model-dependent.

%


{\em Tight-binding models.}
Ab-initio bandstructure calculations for {\sm} \cite{lu_smb6_gutz, smb6_korea_dft, pub6} show that the bands close to the Fermi energy originate from Sm orbitals, namely
in the $d$ shell the $E_g$ quartet ${|d_{x^2-y^2}\uparrow\rangle}$, ${|d_{x^2-y^2}\downarrow\rangle}$, ${|d_{z^2}\uparrow\rangle}$, ${|d_{z^2}\downarrow\rangle}$
and in the $f$ shell, where strong spin-orbit coupling pushes the $j=7/2$ multiplet well above the Fermi energy, the $j=5/2$ multiplet, which is split by cubic crystal field into a $\Gamma_8$ quartet, ${|\Gamma_8^{(1)}\pm\rangle} = {\sqrt\frac{5}{6}|\pm\frac{5}{2}\rangle + \sqrt\frac{1}{6}|\mp\frac{3}{2}\rangle}$,
${|\Gamma_8^{(2)}\pm\rangle} = {|\pm\frac{1}{2}\rangle}$,
and a $\Gamma_7$ doublet,
${|\Gamma_7\pm\rangle} = {\sqrt\frac{1}{6}|\pm\frac{5}{2}\rangle -
\sqrt\frac{5}{6}|\mp\frac{3}{2}\rangle}$.
Any reasonable model must thus consist of a $sc$ lattice of Sm atoms, but the choice of included orbitals and tight-binding parameters (on-site energies $\epsilon$ and hopping parameters $\eta$) is not univocal. In fact, published papers employed models with either 10 bands ($E_g$, $\Gamma_8$, $\Gamma_7$) \cite{prbr_io_smb6}, or 8 bands ($E_g$, $\Gamma_8$) \cite{tki_cubic, prbr_io_smb6},
or 6 bands ($E_g$, $\Gamma_7$) \cite{prbr_io_smb6}, or 4 bands ($s$ doublet, $\Gamma_8^{(2)}$ doublet) \cite{sigrist_tki_prb}. Once the basis orbitals are fixed, different sets of hopping parameters $\eta$ may be used: in Table \ref{table_chern} we summarize some of these choices, with details given in the supplement \cite{suppl_prl}.
%
%
%
%
%

We note that our discussion will be exclusively based on single-particle models: Although the Hubbard repulsion among $f$ electrons is not small, its effect at low temperatures can be captured in terms of renormalized $f$-electron kinetic energy and hybridization \cite{hewson}.
Hence, we can think of working directly with renormalized parameters; our qualitative conclusions are independent of interaction-induced renormalizations.


{\em Distinct topological phases.}
Our main finding is that, upon computing mirror Chern numbers for different {\sm} models, $\C^+_{k_z=0}$ can be either $+2$ or $-2$ and $\C^+_{k_x=k_y}=+1$ or $-1$, while $\C^+_{k_z=\pi}=+1$ always.
This yields a total of four distinct phases, summarized in Table \ref{table_chern}.
In particular, models from Ref.~\onlinecite{tki_cubic} (here denoted as $A^{1z}$) and Ref.~\onlinecite{sigrist_tki_prb} ($sf$) are found to belong to the $(-2,+1,-1)$ phase,
while models from Ref.~\onlinecite{prbr_io_smb6} with $\Gamma_7$ and/or $\Gamma_8$ multiplets realize either the $(+2,+1,-1)$ or the $(-2,+1,+1)$ phase;
the $(+2,+1,+1)$ phase is finally achieved in model $A^{2x}$ obtained by modifying the hybridization term in model $A^{1z}$ from Ref.~\onlinecite{tki_cubic}.


{\em Phases and symmetries of surface states.}
In order to give a transparent physical meaning to the four phases, we relate the mirror Chern numbers to the properties of topological surface states.
First, the absolute value of a mirror Chern number reflects the (minimum) number of Dirac points arising along a high-symmetry line in the 2D surface BZ \cite{kane_bisb,smb6_tci,sigrist_tki_prb}.
For instance, on the (001) surface there are two Dirac points along the $k_{x,y}=0$ directions,
and one along $k_{x,y}=\pi$ and $k_x=\pm k_y$; see Fig. \ref{fig_3dbz}(b).
This is consistent with the three Dirac cones at $\bar\Gamma$, $\bar X$ and $\bar X'$ predicted by parity invariants \cite{takimoto,lu_smb6_gutz,tki_cubic} and observed experimentally \cite{neupane_smb6,mesot_smb6, smb6_arpes_feng,smb6_arpes_reinert,smb6_past_allen}.
Apparently, the absolute values of the $\C$ yield no additional information
for the (001) surface (but they do for the (011) one \cite{smb6_tci}).
This also shows that no other TCI phases are possible, since they would give rise to more Dirac cones than actually observed.

\begin{figure}[tb]
\includegraphics[width=0.44\textwidth]{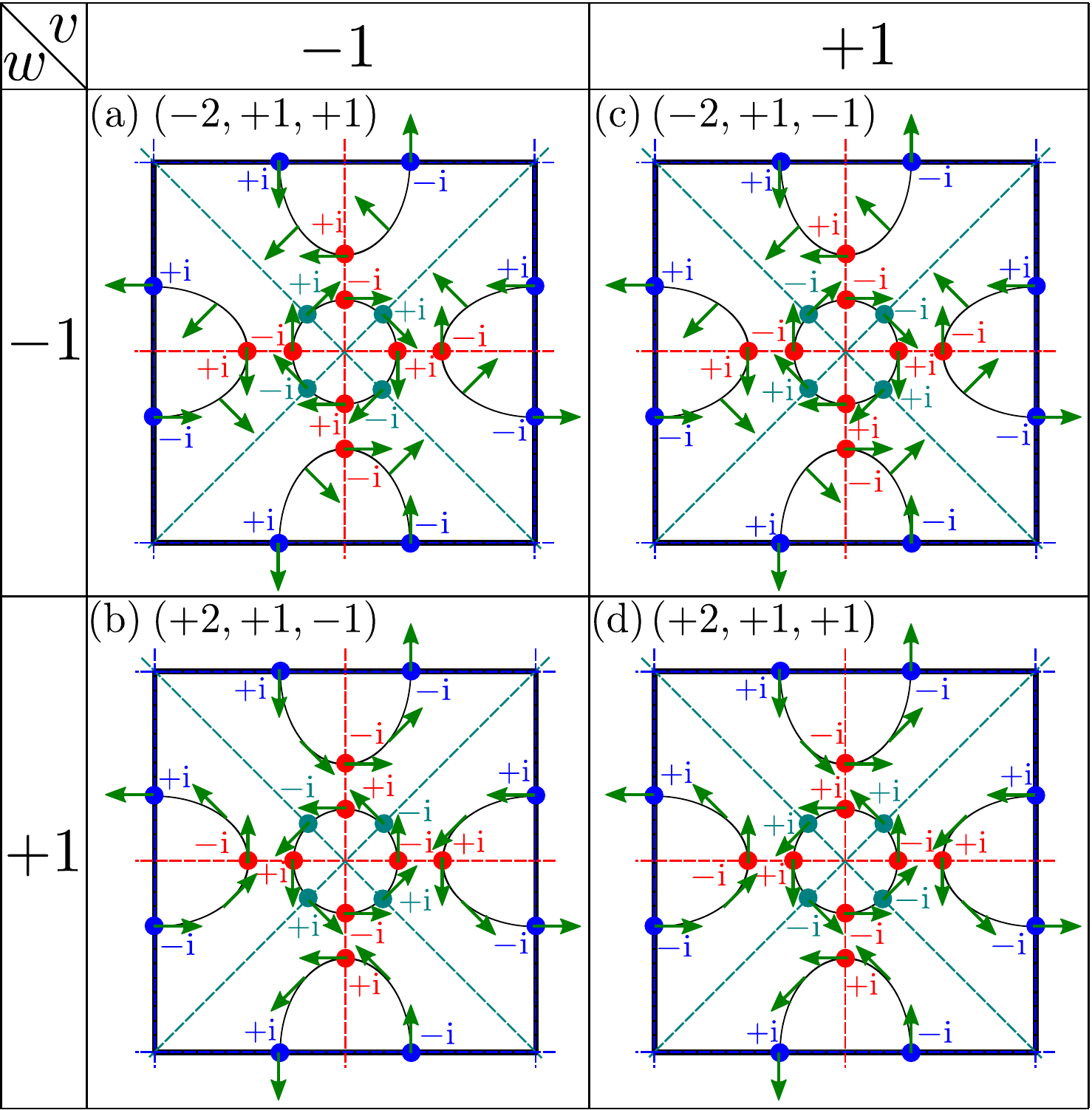}
\caption{Surface-state mirror eigenvalues $\pm\ii$ (defined at surface momenta where the Fermi contour crosses a mirror plane) for the four phases with mirror Chern numbers  $(\C^+_{k_z=0}, \C^+_{k_z=\pi},\C^+_{k_x=k_y})$ as quoted in the panels.
The $\C$ signs can be condensed into $v\equiv\sgn (\C^+_{k_z=0}\C^+_{k_x=k_y})$ and $w\equiv\sgn(\C^+_{k_z=0}\C^+_{k_z=\pi})$, see text.
Green arrows show the surface-state SEV for $\Gamma_8$ states;
for $\Gamma_7$ states (see Table~\ref{table_chern} for models) it would be reversed.
In the FTB model this reversal depends on the relative weight of $\Gamma_7$ and $\Gamma_8$ states, with $\Gamma_8$ dominating for realistic parameters \cite{prbr_io_smb6}.
}
\label{fig_chern}
\end{figure}

However, the sign of a mirror Chern number gives new information, as it fixes the mirror-symmetry eigenvalue of surface states.
%
%
%
For example, $\C^+_{k_z=0}=\C^+_{k_x=0}=+2$ tells that along $k_x=0$ there are two bands with positive velocity along the $\hat{z}\times \hat{x} = \hat{y}$ direction and $M_x$ eigenvalues $+\ii$,
while for $\C^+_{k_z=0}=-2$ these states have eigenvalue $-\ii$ \cite{kane_bisb, sigrist_tki_prb}.
By a repeated use of this property we can assign all the $\pm\ii$ eigenvalues in Fig.~\ref{fig_chern}(a)--(d) for the four possible phases.


{\em Surface-state spin structure.}
Since mirror eigenvalues are not directly measurable,
we now link them to the spin expectation value (SEV) of surface states as observable in SP-ARPES experiments \cite{smb6_arpes_mesot_spin}.
The idea is as follows: If each Dirac-cone state would simply behave as a spin-$1/2$ state with $M_l=C_2(l)=e^{-\ii\pi\sigma_l}= -\ii \sigma_l$, we would have
$M_l|s_l\rangle=-\ii\sigma_l |s_l\rangle=-\ii s_l |s_l\rangle$ ($l=x,y, x\pm y$),
with $|s_l\rangle$ an eigenvector of the Pauli matrix $\sigma_l$ with eigenvalue $s_l=\pm 1$.
Hence, $M_l$ eigenvalues of $\pm \ii$ would immediately give the direction of the spin, e.g.,
an $M_x$ eigenvalue of $-\ii$ would imply a spin pointing towards $+x$.
However, additional minus signs arise for orbitals odd under $M_l$, and, due to the spin-orbit coupling in the $f$ shell, surface states are not eigenstates of spin operators.
By explicitly invoking the properties of the basis states we can derive rules relating the mirror eigenvalues to the SEV:
We find that $M_l|\psi\rangle = -\ii |\psi\rangle$ implies that the SEV for state $|\psi\rangle$ points along direction $\pm l$, with the plus sign always holding
except for $M_{x\pm y}$  acting on $d_{x^2-y^2}$ and $\Gamma_8^{(1)}$ states
and for $M_{x,y}$ on $\Gamma_7$ states \cite{suppl_prl}.
Using these rules,
and the fact that for small momenta around $\bar\Gamma$ the SEV does not depend on the eigenvalues of $M_{x\pm y}$ (see below),
we can construct a full portrait of the SEV for each phase once a basis has been fixed, see Fig.~\ref{fig_chern} for $\Gamma_8$ states.



%
We proceed by analyzing these portraits in detail.
Interpreting the SEV in Fig.~\ref{fig_chern} as true spin \cite{note_true_spin},
the effective low-energy model at the $\bar X$ points can be written as:
\bea
H_{\bar X}&=&v_1 (k_x-\pi)\sigma_y-v_2 k_y\sigma_x,\label{dirac_x1}\\
H_{\bar X'}&=&v_2 k_x \sigma_y-v_1 (k_y-\pi) \sigma_x.\label{dirac_x2}
\eea
Importantly, the velocities $v_{1,2}$ have the same (opposite) sign if $\C^+_{k_z=0}\C^+_{k_z=\pi}>0$ ($<0$), respectively. We introduce $w\equiv\sgn(\C^+_{k_z=0}\C^+_{k_z=\pi})=\sgn(v_1v_2)$ which describes how the mirror eigenvalues and the SEV on the $\bar X$ cones evolve upon $\pi/2$ momentum-space rotations around $\bar X$.
This implies that $w$ is a {\em winding number}: when $w=+1$ spin and momentum rotate around $\bar X$ in the same direction, Figs.~\ref{fig_chern}(b,d), while for $w=-1$ they rotate in the opposite direction, Figs.~\ref{fig_chern}(a,c). We note that $w=-1$ is only allowed for cones at low-symmetry points, see below.

Similarly, we introduce $v\equiv\sgn (C^+_{k_z=0}C^+_{k_x=k_y})$ which now describes how the mirror eigenvalues on the $\bar\Gamma$ cone evolve upon $\pi/4$ rotations. It turns out, however, that $v$ does not affect the SEV of the $\bar\Gamma$ cone: For sufficiently small momenta the system must display continuous rotation symmetry around $\bar\Gamma$ which implies a fixed angle between SEV and momentum along the contour (i.e., a winding number $+1$).
Thus the effective low-energy model at the $\bar \Gamma$ point is independent of $v$, and reads
\be
H_{\bar\Gamma}=v_0w (k_x\sigma_y- k_y\sigma_x)~~(v_0>0), 
\ee
with spin and momentum always forming a mutual angle $\pm\pi/2$ given by the sign of $v_0w$, i.e., the chirality.
We note that the $w$ factor in $H_{\bar\Gamma}$ is needed to give the correct relative spin direction along $k_x$ and $k_y$ between the $\bar\Gamma$ and the $\bar X$ cones, which is fixed by $|\C^+_{k_z=0}|=2$.
%
%
Using $H_{\bar\Gamma}$ we can show \cite{suppl_prl} that $v$ instead dictates symmetry properties of the states composing the $\bar\Gamma$ cone for small momenta:
for $v=+1$ these are $d_{z^2}$ and $\Gamma_8^{(2)}$ ($X_6$ symmetry representation) whereas for $v=-1$ these are $d_{x^2-y^2}$, $\Gamma_8^{(1)}$ and $\Gamma_7$ (corresponding to $X_7$).


{\em Relation between $v$,$w$ and electronic structure.}
%
In order to determine the TCI phase of {\sm} we now connect the parameters $v$ and $w$ to the bulk bandstructure. Surface states near $\bar\Gamma$ can be computed by perturbatively expanding the Hamiltonian around $X=(0,0,\pi)$  \cite{liu_ti_model,dzero_pert},
where it decouples into a subspace with symmetry representation $X_7$,
composed from $d_{x^2-y^2}$, $\Gamma_8^{(1)}$, and $\Gamma_7$ states,
and a subspace with symmetry representation $X_6$ and $d_{z^2}$, $\Gamma_8^{(2)}$ states.
Band inversion can be achieved in only one of these subspaces:
in model $A^{1z}$ from Refs.~\onlinecite{tki_cubic,dzero_pert}
it is assumed to be in the subspace $X_6$,
while DFT calculations \cite{smb6_korea_dft} display band inversion in subspace $X_7$.
%
%
Since surface states at $\bar\Gamma$ only exist in the subspace where band inversion is achieved \cite{dzero_pert},
and this depends on 
which $d$ band has a minimum at $X$,
we can link these two options to the symmetry of the $d$-band minimum at $X$:
$d_{z^2}$ ($X_6^+$) in the first case, leading to $v=+1$,
$d_{x^2-y^2}$ ($X_7^+$) in the second, leading to $v=-1$.
%
Since
we expect DFT to be reliable for weakly correlated orbitals,
we conclude that $v=-1$ in real {\sm}.

Once the band-inversion subspace, and thus $v$, is fixed by the choice of the hopping terms, $w$ depends on the hybridization term: out of those which lead to a full gap, some lead to $w=+1$, some others to $w=-1$, as shown in Table~\ref{table_chern}.
When considering $\Gamma_8$ states, DFT calculations \cite{pub6,prbr_io_smb6} show the largest of the hybridization terms to be $\eta_z^{v1}$ \cite{model_parameters},
which, however, alone does not lead to a gap for $v=-1$, since it does not couple the inverted bands $d_{x^2-y^2}$ and $\Gamma_8^{(1)}$ along $\Gamma$--$X$ for symmetry reasons.
The second-largest term is $\eta_z^{v2}$, which gives $w=+1$ ($\Gamma_8^{2z}$ in Table \ref{table_chern}).
However, a competition with the $w=-1$ phase can be observed when retaining both $\Gamma_7$ and $\Gamma_8$ multiplets \cite{prbr_io_smb6},
since the $\Gamma_7$ doublet alone with $E_g$ is in the $w=-1$ phase, $\eta_7^{v2}$ being the largest hybridization term in this case ($\Gamma_7^{2z}$).


{\em Choice of phase and minimal model.}
We now make use of the following ingredients:
(i) The band inversion properties from DFT calculations yield $v=-1$.
(ii) The SP-ARPES experiment of Ref.~\onlinecite{smb6_arpes_mesot_spin} shows a winding number on $\bar X$ cones of $w=+1$.
Together, this uniquely yields $(+2,+1,-1)$ as the best candidate TCI phase for {\sm}.


%
At the same time, we realize that the $\Gamma_8^{2z}$ model, obtained retaining the $E_g$ and $\Gamma_8$ quartets, with $\eta_z^{d2}/\eta_z^{d1}\sim\eta_z^{f2}/\eta_z^{f1}\sim -3/8$ and
the $\eta_z^{v2}$ hybridization term, is the simplest one giving the correct mirror Chern numbers $(+2,+1,-1)$
and reproducing as closely as possible the DFT bandstructure. We propose it as a new minimal model for {\sm}, with a concrete tight-binding parametrization given in the supplement \cite{suppl_prl}.
We stress that this conclusion is based on the experimental result $w=+1$; $w=-1$ would instead lead to a minimal model entailing the $E_g$ quartet and the $\Gamma_7$ doublet ($\Gamma_7^{2z}$).
The competition between the $w=+1$ and $w=-1$ phases, emerging from the $\Gamma_8$ quartet and the $\Gamma_7$ doublet, respectively,
depends on model details which have not been obtained with sufficient accuracy from DFT.
This applies for instance to the $\Gamma_7$--$\Gamma_8$ energy difference whose variation can drive a topological phase transition between states with different $w$.

\begin{figure}[tb]
\includegraphics[width=0.49\textwidth]{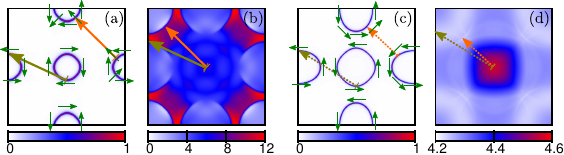}
\caption{
Isoenergy contours with SEV (a,c) and QPI signal (b,d) at the Fermi level for two different {\sm} tight-binding models. Large arrows show intercone scattering wavevectors.
(a,b): Four-band model $sf$ in the $(-2,+1,-1)$ phase ($w=-1$);
(c,d): Eight-band model $\Gamma_8^{2z}$ in the $(+2,+1,-1)$ phase ($w=+1$);
with parameters given in the supplement \cite{suppl_prl}.
An intercone QPI signal is seen in (b), but not in (d), dictated by the surface-state spin structure.
}\label{fig_qpi_2}
\end{figure}

{\em Quasiparticle interference (QPI).}
QPI can be used as a probe of the topological character of surface states \cite{xue,yazdani,guo_franz_sti}, owing to their spin texture;
in TCIs, impurities which do not break mirror symmetries cannot induce transitions between states with opposite mirror eigenvalues \cite{tci_qpi_theory}.
%
%
%
%
Here we show how the winding number $w$ affects intercone scattering.
Comparing the spin portraits for two relevant models in Figs.~\ref{fig_qpi_2}(a) and (c) we see that, for pairs of stationary points \cite{liu_qpi_spa} (states which have the same tangent to the isoenergy contour) belonging to different $\bar X$ cones, $w=-1$ ($w=+1$) implies roughly parallel (antiparallel) spin, respectively.
Consequently, $w=-1$ yields a QPI peak from intercone scattering \cite{prb_io_tki,yu_smb6_qpi}, Fig.~\ref{fig_qpi_2}(b), while $w=+1$ does not \cite{prbr_io_smb6}, Fig.~\ref{fig_qpi_2}(d).
Thus, QPI can probe $w$ by looking at intercone scattering, 
and hence provides information on the spin structure of the Dirac cones and, indirectly, on mirror Chern numbers.

{\em Conclusions.}
In this Letter we have studied mirror Chern numbers in different models for the correlated topological insulator {\sm}.
%
The absolute values of the mirror Chern numbers are fixed by symmetry properties of the atomic orbitals and the bandstructure and are thus independent of model details,
while their signs are model-dependent.
This yields four possible phases -- all sharing the $\mathbb{Z}_2$ topological indices (1,111) --
which can be characterized by the two combinations of signs $v\equiv\sgn (\C^+_{k_z=0}\C^+_{k_x=k_y})$ and $w\equiv\sgn(\C^+_{k_z=0}\C^+_{k_z=\pi})$.
These have a straightforward physical interpretation in terms of the symmetry of the $d$-band minimum at $X$ and the spin texture of the $\bar X$ cone on a (001) surface, respectively.
%
Our analysis constrains {\sm} to be in the $(\C^+_{k_z=0},\C^+_{k_z=\pi},\C^+_{k_x=k_y})=(+2,+1,-1)$ phase, by using the results of DFT calculations
%
and of SP-ARPES \cite{smb6_arpes_mesot_spin} which give $v=-1$ and $w=+1$, respectively.
For the $w=+1$ case we predict a weak QPI signal from intercone scattering.
%

From our analysis we propose a minimal tight-binding model for {\sm} consisting of 8 orbitals (the $E_g$ and $\Gamma_8$ quartets). The model equals the one of Ref.~\onlinecite{tki_cubic} in the choice of the orbital basis but differs in both the kinetic energy and the hybridization parameters; it better reproduces DFT and SP-ARPES data, at the price of introducing second-neighbor hopping parameters.
%
We finally speculate that chemical substitution can tune both the $\Gamma_7$--$\Gamma_8$ energy difference and the hopping parameters, possibly leading to a topological phase transition from the $w=+1$ to the $w=-1$ state. Such a transition is accompanied by the closing of the bulk gap.

We believe that our results advance the understanding of the complex TKI material {\sm} (and its cousin {\pu} \cite{pub6}), and potentially of other TIs, too.

\acknowledgments

This research was supported by the DFG through SFB 1143 and GRK 1621
as well as by the Helmholtz association through VI-521.

{\it Note added:}
At the day of submission of this paper, a related paper \cite{legner15} appeared on arXiv, which reaches conclusions compatible with ours. For comparison we note that their Chern number $C_d$ equals our $(-\C^+_{k_x=k_y})$.

\bibliographystyle{apsrev4-1}
\bibliography{tki}

\begin{thebibliography}{38}
\bibitem [{mod()}]{model_parameters}%
  \BibitemOpen
  \href@noop {} {}\bibinfo {note} {We denote the $\eta$ parameters as follows:
  the superscript denotes the $d$, $s$ or $f$ shell for the kinetic energy, and
  $v$ for the hybridization, and the layer of NN; the subscript, if it involves
  $d_{x^2-y^2}$ or $\Gamma_8^{(1)}$ along the 001 or 110 direction, is $x$, if
  it involves $d_{z^2}$ or $\Gamma_8^{(2)}$ along the 100 or 011 direction, is
  $z$, if it involves $\Gamma_7$ is $7$. So, for example, $\eta_z^{v1}$ is the
  1NN hybridization term along 001 between $d_{z^2}$ and
  $\Gamma_8^{(2)}$.}\BibitemShut {Stop}%
\bibitem [{not()}]{note_true_spin}%
  \BibitemOpen
  \href@noop {} {}\bibinfo {note} {In a microscopic quantitative
  treatment\cite{liu_ti_model}, Pauli matrices will rather describe a Kramer's
  pair $|\psi_+\rangle$, $|\psi_-\rangle$ whose SEV is such that
  $|\psi_+\rangle \sim |\uparrow\rangle$, $|\psi_-\rangle \sim
  |\downarrow\rangle$, but our qualitative conclusions will not
  change}\BibitemShut {NoStop}%
\end{thebibliography}

\end{document}


\title{
Supplemental material for ``Distinct topological crystalline phases in models for the strongly correlated topological insulator {\sm}''
}

\author{Pier Paolo Baruselli}
\author{Matthias Vojta}
\affiliation{Institut f\"ur Theoretische Physik, Technische Universit\"at Dresden, 01062 Dresden, Germany}


\maketitle


\section{Model Hamiltonians}

In this section we provide explicit expressions for the tight-binding Hamiltonians referred to in the main text and Table~1.

\subsection{Parametrization}

We will use $\sigma$, $\tau$, and $\mu$ Pauli matrices as follows:
$\sigma$ matrices act in the (pseudo)spin subspace,
$\tau$ in the orbital subspace, and $\mu$ in the $d$-$f$ subspace.
Then, terms proportional to $\mu_0$ and $\mu_z$ denote the kinetic energy (with subscript $0$ representing the $2\times2$ unit matrix) while terms proportional to $\mu_y$ correspond to $d$-$f$ hybridization.
For convenience, we introduce $\mu_1=(\mu_0+\mu_z)/2$, $\mu_2=(\mu_0-\mu_z)/2$, $\tau_1=(\tau_0+\tau_z)/2$, $\tau_2=(\tau_0-\tau_z)/2$:
$\mu_1$ projects onto the $d$ subspace, $\mu_2$ onto the $f$ subspace, $\tau_1$ onto the orbital subspace 1 ($d_{x^2-y^2}$, $\Gamma_8^{(1)}$), and $\tau_2$ onto the orbital subspace 2 ($d_{z^2}$, $\Gamma_8^{(2)}$).
We also employ abbreviations $c_x=\cos k_x$, $c_y=\cos k_y$, $c_z=\cos k_z$, $s_x=\sin k_x$, $s_y=\sin k_y$, $s_z=\sin k_z$.

\bigskip

We consider three classes of models:

\begin{widetext}
\begin{itemize}
\item 8-band models consisting of the $E_g(d)$ quartet interacting with the $\Gamma_8(f)$ quartet.\\The parameters are
${\epsilon^{d}=\langle d_{x^2-y^2} |H|d_{x^2-y2}\rangle_{000}}=\langle d_{z^2} |H|d_{z^2}\rangle_{000}$,
${\epsilon_8^{f}=\langle \Gamma_8^{(1)}+|H|\Gamma_8^{(1)}+\rangle_{000}}=\langle \Gamma_8^{(2)}+|H|\Gamma_8^{(2)}+\rangle_{000}$,
${\eta_x^{d1}=\langle d_{x^2-y^2} |H|d_{x^2-y2}\rangle_{001}}$,
${\eta_z^{d1}=\langle d_{z^2}|H|d_{z^2}\rangle_{001}}$,
${\eta_z^{d2}=\langle d_{z^2}|H|d_{z^2}\rangle_{110}}$,
${\eta_x^{f1}=\langle \Gamma_8^{(1)}+|H|\Gamma_8^{(1)}+\rangle_{001}}$,
${\eta_z^{f1}=\langle \Gamma_8^{(2)}+|H|\Gamma_8^{(2)}+\rangle_{001}}$,
${\eta_z^{f2}=\langle \Gamma_8^{(2)}+|H|\Gamma_8^{(2)}+\rangle_{110}}$,
${\eta_x^{v1}=\langle d_{x^2-y^2}\uparrow|H|\Gamma_8^{(1)}+\rangle_{001}}$,
${\eta_z^{v1}=\langle d_{z^2}\uparrow|H|\Gamma_8^{(2)}+\rangle_{001}}$,
${\eta_x^{v2}=\langle d_{x^2-y^2}\uparrow|H|\Gamma_8^{(1)}+\rangle_{110}}$,
${\eta_z^{v2}=\langle d_{z^2}\uparrow|H|\Gamma_8^{(2)}+\rangle_{110}}$,
and Hamiltonian $H=H_{E_g}+H_{\Gamma_8}+V_{E_g-\Gamma_8}$:
\bea
H_{i}(\bk)&=&\epsilon^i\tau_0\sigma_0\mu_i-t_i\sigma_0\mu_i\cdot\nonumber\\
&&\cdot\{[(c_x+c_y+c_z)(\eta_z^{i1}+\eta_x^{i1})+2\eta_z^{i2}(c_xc_y+c_yc_z+c_zc_x)]\tau_0+\nonumber\\
&&+[(c_x+c_y-2c_z)(\eta_z^{i1}-\eta_x^{i1})+2\eta_z^{i2}(-2c_xc_y+c_yc_z+c_zc_x)]\tau_z/2+\nonumber\\
&&+[\sqrt{3}(c_x-c_y)(\eta_x^{i1}-\eta_z^{i1}+2c_z\eta_z^{i2})]\tau_x/2\},
\eea
with $i=E_g(d)\equiv 1,\Gamma_8(f)\equiv 2$,
\bea
V_{E_g-\Gamma_8}(\bk)&=&-\mu_y V \{s_x\sigma_x\tau_0[\eta_x^{v1}+\eta_z^{v1}+2(c_z+c_y)(\eta_z^{v2}+\eta_x^{v2})]+\nonumber\\
&&+s_y\sigma_y\tau_0[\eta_x^{v1}+\eta_z^{v1}+2(c_z+c_x)(\eta_z^{v2}+\eta_x^{v2})]+\nonumber\\
&&+s_z\sigma_z\tau_0[\eta_x^{v1}+\eta_z^{v1}+2(c_x+c_y)(\eta_z^{v2}+\eta_x^{v2})]+\nonumber\\
&&+s_x\sigma_x\tau_z[-\eta_x^{v1}+\eta_z^{v1}+2(c_z-2c_y)(\eta_z^{v2}-\eta_x^{v2})]/2+\nonumber\\
&&+s_y\sigma_y\tau_z[-\eta_x^{v1}+\eta_z^{v1}+2(c_z-2c_x)(\eta_z^{v2}-\eta_x^{v2})]/2+\nonumber\\
&&+s_z\sigma_z\tau_z[\eta_x^{v1}-\eta_z^{v1}+(c_x+c_y)(\eta_z^{v2}-\eta_x^{v2})]+\nonumber\\
&&+\sqrt{3}s_x\sigma_x\tau_x[\eta_x^{v1}-\eta_z^{v1}+2c_z(\eta_z^{v2}-\eta_x^{v2})]/2+\nonumber\\
&&+\sqrt{3}s_y\sigma_y\tau_x[-\eta_x^{v1}+\eta_z^{v1}-2c_z(\eta_z^{v2}-\eta_x^{v2})]/2\}+\nonumber\\
&&+\sqrt{3}s_z\sigma_z\tau_x(c_x-c_y)(\eta_z^{v2}-\eta_x^{v2})\};
\eea

\item 6-band models consisting of the $E_g(d)$ quartet interacting with the $\Gamma_7(f)$ doublet.\\Here the parameters are ${\epsilon^{d}}$,
${\epsilon_7^{f}=\langle \Gamma_7+|H|\Gamma_7+\rangle_{000}}$,
$\eta_x^{d1}$, $\eta_z^{d1}$, $\eta_z^{d2}$,
${\eta_7^{f1}=\langle \Gamma_7+|H|\Gamma_7+\rangle_{001}}$,
${\eta_7^{f2}=\langle \Gamma_7+|H|\Gamma_7+\rangle_{110}}$,
${\eta_7^{f3}=\langle \Gamma_7+|H|\Gamma_7+\rangle_{111}}$,
${\eta_7^{v1}=\langle d_{x^2-y^2}\uparrow|H|\Gamma_7+\rangle_{001}}$,
${\eta_7^{v2}=\langle d_{z^2}\uparrow|H|\Gamma_7+\rangle_{110}}$;
and Hamiltonian $H=H_{E_g}+H_{\Gamma_7}+V_{E_g-\Gamma_7}$:
\bea
H_{\Gamma_7}(\bk)&=&-t_{\Gamma_7}\sigma_0\mu_2[2\eta_7^{f1}(c_x+c_y+c_z)+\nonumber\\
&&+4\eta_7^{f2}(c_xc_y+c_yc_z+c_zc_x)+8\eta_7^{f3}c_xc_yc_z],
\eea
\bea
V_{E_g-\Gamma_7}(\bk)&=&-\mu_y V\{2s_z\sigma_z\omega_0[(\eta_7^{v1}-\sqrt{3}(c_x+c_y)\eta_7^{v2}]+\nonumber\\
&&+(s_x\sigma_x+s_y\sigma_y)\omega_0[-\eta_7^{v1}+2\sqrt{3}c_z\eta_7^{v2}]+\nonumber\\
&&+2s_z\sigma_z\omega_1(c_y-c_x)\eta_7^{v2}+\nonumber\\
&&+s_x\sigma_x\omega_1[-\sqrt{3}\eta_7^{v1}+2\eta_7^{v2}(c_z+2c_y)]+\nonumber\\
&&+s_y\sigma_y\omega_1[\sqrt{3}\eta_7^{v1}-2\eta_7^{v2}(c_z+2c_x)]\},
\eea
where $\omega_0=(1,0)$ and $\omega_1=(0,1)$ act in the orbital space, projecting respectively onto the $d_{x^2-y^2}$ and $d_{z^2}$ doublets,
and $H_{E_g}(\bk)$ is the same as from the 8-band model;

\item 4-band model consisting of conduction $(s)$ electrons interacting with the $\Gamma_8^{(2)}(f)$ doublet.\\ This has parameters
${\epsilon^{s}=\langle s |H|s\rangle_{000}}$,
${\epsilon^{f}=\langle \Gamma_8^{(2)}+|H|\Gamma_8^{(2)}+\rangle_{000}}$,
${\eta^{s1}=\langle s|H|s\rangle_{001}}$,
${\eta^{s2}=\langle s|H|s\rangle_{110}}$,
${\eta^{f1}=\langle \Gamma_8^{(2)}+|H|\Gamma_8^{(2)}+\rangle_{001}}$,
${\eta^{f2}=\langle \Gamma_8^{(2)}+|H|\Gamma_8^{(2)}+\rangle_{110}}$,
${\eta^{v1}=\langle s\uparrow|H|\Gamma_8^{(2)}+\rangle_{001}}$:
and Hamiltonian $H=H_s+H_{\Gamma_8^{(2)}}+V_{s-\Gamma_8^{(2)}}$:
\bea
H_{i}(\bk)&=&-2t_{i}\mu_i [\eta^{i1}(c_x+c_y+c_z)+2\eta^{i2}(c_xc_y+c_yc_z+c_zc_x)]\sigma_0,\\
V_{s-\Gamma_8^{(2)}}(\bk)&=&2V\mu_y\eta^{v1}(s_x\sigma_x+s_y\sigma_y+s_z\sigma_z),\label{v_sf}
\eea
with $i=s\equiv 1,\Gamma_8^{(2)}(f)\equiv 2$.
\end{itemize}

\end{widetext}

We now relate these general model Hamiltonians to models which have appeared in the literature and to those which are listed in Table~1 of the main text.

\begin{itemize}

\item The model by Legner \textit{et al.}\cite{sigrist_tki_prb} (here abbreviated as $sf$) is identical to the 4-band model, setting $\eta^{s2}/\eta^{s1}=\eta^{f2}/\eta^{f1}\sim -0.5$.

\item The 8-band model by Alexandrov \textit{et al.}\cite{tki_cubic} ($A^{1z}$) is obtained in the $E_g$-$\Gamma_8$ basis by setting $\eta_z^{d1}/\eta_x^{d1}\sim \eta_z^{f1}/\eta_x^{f1}\sim -0.3$, and keeping the $\eta_z^{v1}$ 1NN term in the hybridization. For completeness, we have also considered the same model with the $\eta_x^{v2}$ ($A^{2x}$)  2NN term in the hybridization, since it belongs to a different phase.

\item Our $\Gamma_8^{1x}$ and $\Gamma_8^{2z}$ 8-band models are obtained retaining the $E_g$ and the $\Gamma_8$ quartets, with $\eta_z^{d2}/\eta_z^{d1}\sim -3/8$, $\eta_z^{f2}/\eta_z^{f1}\sim -1/2$ and keeping either the 1NN $\eta_x^{v1}$ ($\Gamma_8^{1x}$) or the 2NN $\eta_z^{v2}$ ($\Gamma_8^{2z}$) term in the hybridization.

\item Our $\Gamma_7^{1x}$ and $\Gamma_7^{2z}$ 6-band models are obtained retaining the $E_g$ quartet and the $\Gamma_7$ doublet, with $\eta_z^{d2}/\eta_z^{d1}\sim-3/8$, $\eta_7^{f3}/\eta_7^{f2}\sim 1/2$, and keeping either the 1NN $\eta_7^{v1}$ ($\Gamma_7^{1x}$) or the 2NN $\eta_7^{v2}$ ($\Gamma_7^{2z}$) term in the hybridization.

\item The full tight-binding calculation ($FTB$) of Ref. \onlinecite{prbr_io_smb6} comprises the $E_g$ and the $\Gamma_8$ quartets, and the $\Gamma_7$ doublet, for a total of 10 orbitals per site. Parameters are retained up to 7NN, such that we cannot give a complete analytic expression.

\item The model by Takimoto \cite{takimoto} ($T$) is equivalent to our $E_g-\Gamma_8$ model, retaining, however, in addition to more $f$ terms, also the $\eta_z^{v1}$ hybridization term. This term alone, in our approach, does not lead to an insulator, since it does not provide a gap along the $\Gamma-X$ direction in the BZ. Therefore we have not discussed this model in the paper.

\end{itemize}

\subsection{Realistic tight-binding parameters}

A plausible parameter set for the $\Gamma_8^{2z}$ model, which essentially reproduces the DFT bandstructure of {\pu},\cite{pub6} is:\cite{prbr_io_smb6}
$t_d=1$\,eV, $t_f=0.01$\,eV,
$\epsilon^f_8-\epsilon_d=-2$\,eV,
$\eta_z^{d1}=0.8$, $\eta_z^{d2}=-0.3$,
$\eta_z^{f1}=-4$, $\eta_z^{f2}=2$,
$V\eta_z^{v2}=0.06$\,eV.
Similarly, a plausible parameter set for the $\Gamma_7^{2z}$ model is:
$t_d=1$\,eV, $t_f=0.01$\,eV,
$\epsilon^f_7-\epsilon_d=-1.9$\,eV,
$\eta_z^{d1}=0.8$, $\eta_z^{d2}=-0.3$,
$\eta_7^{f2}=2.5$, $\eta_z^{f3}=1.25$,
$V\eta_7^{v2}=0.05$\,eV.
The effect of electronic correlations, mainly absent from DFT, will be a downward renormalization of $t_f$ and $V$, see Ref.~\onlinecite{prbr_io_smb6} for a more extensive discussion.

Panels (a) and (b) of Fig. 3 of the main text have been calculated using the $sf$ model with
$t_s=1$eV, $t_f=0.003$eV,
$\epsilon^f-\epsilon_2=-2$\,eV,
$\eta^{s1}=\eta^{f1}=1$, $\eta^{s2}=\eta^{f2}=-0.5$,
$V\eta^{v1}=0.2$\,eV.
Panels (c) and (d) of the same figure have been obtained using the $\Gamma_8^{2z}$ model with
$t_d=1$\,eV, $t_f=0.01$\,eV,
$\epsilon^f_8-\epsilon_d=-2$\,eV,
$\eta_z^{d1}=0.8$, $\eta_z^{d2}=-0.3$,
$\eta_z^{f1}=-4$, $\eta_z^{f2}=2$,
$V\eta_z^{v2}=0.02$\,eV, $V\eta_z^{v1}=-0.06$\,eV.
Compared to the parameter set from above, this includes the additional hybridization parameter $\eta_z^{v1}$ which improves the quantitative agreement with DFT, but does not change the topological properties.

\begin{table*}
$$
\begin{array}{|c|cccc|cc|}\hline
(\sigma^x,\sigma^y, \sigma^z)&\Gamma_8^{(1)}+ & \Gamma_8^{(1)}- & \Gamma_8^{(2)}+ & \Gamma_8^{(2)}- & \Gamma_7 +& \Gamma_7 -\\\hline
\Gamma_8^{(1)}+&\frac{11}{21}(0,0,-1) & \frac{5}{21}(-1,\ii,0) & (0,0,0) & \frac{2\sqrt{3}}{21}(-1,-\ii,0) & \frac{4\sqrt{5}}{21}(0,0-1)&\frac{{2\sqrt{5}}}{21} (1,-\ii,0)\\
\Gamma_8^{(1)}-&\frac{5}{21}(-1,-\ii,0) & \frac{11}{21} (0,0,1) & \frac{2\sqrt{3}}{21} (-1,\ii,0) & (0,0,0)&\frac{2\sqrt{5}}{21}(1,\ii,0)&\frac{4\sqrt{5}}{21}(0,0,1)\\
\Gamma_8^{(2)}+&(0,0,0) & \frac{2\sqrt{3}}{21}(-1,-\ii,0) & \frac{1}{7} (0,0,-1) & \frac{3}{7}(-1,\ii,0)& (0,0,0)& \frac{{2}\sqrt{15}}{21}(1,\ii,0)\\
\Gamma_8^{(2)}-&\frac{{2}\sqrt{3}}{21}(-1,\ii,0)&(0,0,0)&\frac{3}{7} (-1,-\ii,0)&\frac{1}{7} (0,0,1) &\frac{{2}\sqrt{15}}{21}(1,-\ii,0)&(0,0,0)\\\hline
\Gamma_7+&\frac{4\sqrt{5}}{21}(0,0,-1)&\frac{2\sqrt{5}}{21} (1,-\ii,0) &(0,0,0)&\frac{{2}\sqrt{15}}{21}(1,\ii,0)&\frac{5}{21}(0,0,1)&\frac{5}{21}(1,-\ii,0)\\
\Gamma_7-&\frac{2\sqrt{5}}{21}(1,\ii,0)&\frac{4\sqrt{5}}{21}(0,0,1)&\frac{{2}\sqrt{15}}{21}(1,-\ii,0)&(0,0,0)&\frac{5}{21}(1,\ii,0)& \frac{5}{21}(0,0,-1)\\\hline
  \end{array}
$$
\caption{SEV in the $\Gamma_7-\Gamma_8$ basis.}\label{table_spin}
\end{table*}

\section{SEV from mirror symmetry eigenvalues}

Here we show how to compute the spin expectation value (SEV)  of surface states from the corresponding mirror eigenvalues.
Mirror operators are defined as $M_l\equiv M_l^{orb}M_l^{spin}$, where
$M_l^{orb}:l\rightarrow -l,$ ($l=x,y,z$),
$M_{x\pm y}^{orb}:x\leftrightarrow \mp y$,
and
$M_l^{spin}=e^{-\ii\pi\sigma_l}=-\ii \sigma_l$ ($l=x,y,z$),
$M_{x\pm y}^{spin}=-\ii(\sigma_x\pm \sigma_y)/{\sqrt{2}} \equiv-\ii\sigma_{x\pm y}$.

We now determine the action of these mirror operators on $E_g$, $\Gamma_7$, $\Gamma_8$ states:
\bea
M_l |d_m \sigma \rangle&=&-\ii \sigma_l |d_m \sigma \rangle, ~ m=x^2-y^2,z^2;\sigma=\uparrow, \downarrow,~~\\
M_l |\Gamma_m \sigma \rangle&=&+ \ii\sigma_l |\Gamma_m \sigma \rangle, ~ m=8^{(1)}, 8^{(2)}, 7;\sigma=\pm,~~
\eea
where $l=x,y,z$, and $ \sigma_l$ acts on the subspace spanned by $\uparrow,\downarrow$ for $d$ states, or by $+,-$ for $f$ states; moreover
\bea
M_{x \pm y}|d_{x^2-y^2} \sigma \rangle&=&+\ii\frac{ \sigma_x\pm \sigma_y}{\sqrt{2}}|d_{x^2-y^2} \sigma \rangle,\hspace{10pt}\\
M_{x \pm y}|d_{z^2} \sigma \rangle&=&-\ii\frac{ \sigma_x\pm \sigma_y}{\sqrt{2}}|d_{z^2} \sigma \rangle,\\
M_{x \pm y}|\Gamma_m \sigma \rangle&=&-\ii\frac{ \sigma_x\pm \sigma_y}{\sqrt{2}}|\Gamma_m \sigma \rangle,\hspace{5pt}m=8^{(1)},7,~~~~~\\
M_{x \pm y}|\Gamma_m \sigma \rangle&=&+\ii\frac{ \sigma_x\pm \sigma_y}{\sqrt{2}}|\Gamma_m \sigma \rangle,\hspace{5pt}m=8^{(2)}.
\eea

We can now find which states belong to eigenvalues $(\pm \ii)$.
We start with $M_x$.
States with eigenvalue $(-\ii)$ are:
\bea
|\psi^{-}_{1,x}\rangle&=&|d_{x^2-y^2}\rangle (1,1)/\sqrt{2},\\
|\psi^{-}_{2,x}\rangle&=&|d_{z^2}\rangle (1,1)/\sqrt{2},\\
|\psi^{-}_{3,x}\rangle&=&|\Gamma_8^{(1)}\rangle (1,-1)/\sqrt{2},\\
|\psi^{-}_{4,x}\rangle&=&|\Gamma_8^{(2)}\rangle (1,-1)/\sqrt{2},\\
|\psi^{-}_{5,x}\rangle&=&|\Gamma_7\rangle (1,-1)/\sqrt{2}
\eea
where the two-dimensional vector is in the ($\uparrow, \downarrow$) ($d$ states) or ($+,-$) ($f$ states) subspace.
The states $|\psi^{-}_{1,x}\rangle$ and $|\psi^{-}_{2,x}\rangle$ are spin eigenstates, pointing towards the positive $x$ ($+x$) direction, but the other three states are not, because $+$ and $-$ are not spin indices.
We can, however, compute the SEV on these states with the use of Table~\ref{table_spin}.
Now, on $\Gamma_8$ states, $\langle \sigma_x,\sigma_y,\sigma_z\rangle \sim (-\hat s_x,-\hat s_y,-\hat s_z)$,
where on the left-hand side $\sigma_i$ are true spin operators, while on the right-hand side $\hat s_i$ act on the pseudospin space;
for $\Gamma_7$ states, instead, $\langle \sigma_x,\sigma_y,\sigma_z\rangle \sim (+\hat s_x,+\hat s_y,+\hat s_z)$.
So, for states $|\psi^{-}_{3,x}\rangle$ and $|\psi^{-}_{4,x}\rangle$ the SEV is pointing in the $+x$ direction,
while for $|\psi^{-}_{5,x}\rangle$ it is pointing in the $-x$ direction.

To conclude, for a state $|\psi\rangle$ with $M_x|\psi\rangle=-\ii|\psi\rangle$, even though it is not in general an eigenstate of the spin,
its SEV is pointing along $+x$ if it is a $d$ or a $\Gamma_8$ state, but along $-x$ if it is a $\Gamma_7$ state.
The opposite holds for eigenvalue $+\ii$.
We can repeat the same reasoning for $M_y$, and conclude that, for a state $|\psi\rangle$ with $M_y|\psi\rangle=-\ii|\psi\rangle$ its SEV is pointing along $+y$ if it is a $d$ or a $\Gamma_8$ state, but along $-y$ if it is a $\Gamma_7$ state.

For $M_{x-y}$ the eigenstates with eigenvalue $(-\ii)$ are:
\bea
|\psi^{-}_{1,x-y}\rangle&=&|d_{x^2-y^2}\rangle (1,\e^{3\ii\pi/4})/\sqrt{2},\\
|\psi^{-}_{2,x-y}\rangle&=&|d_{z^2}\rangle (1,\e^{-\ii\pi/4})/\sqrt{2},\\
|\psi^{-}_{3,x-y}\rangle&=&|\Gamma_8^{(1)}\rangle (1,\e^{-\ii\pi/4})/\sqrt{2},\\
|\psi^{-}_{4,x-y}\rangle&=&|\Gamma_8^{(2)}\rangle (1,\e^{3\ii\pi/4})/\sqrt{2},\\
|\psi^{-}_{5,x-y}\rangle&=&|\Gamma_7\rangle (1,\e^{-\ii\pi/4})/\sqrt{2},
\eea
which means that $|\psi^{-}_{1,x-y}\rangle$ is a spin eigenstate with spin pointing along $(- x+ y)$, while $|\psi^{-}_{2,x-y}\rangle$ points along  $(x-y)$;
for $|\psi^{-}_{3,x-y}\rangle$ the SEV points along $(- x+ y)$, for $|\psi^{-}_{4,x-y}\rangle$ along $(x- y)$, and for
$|\psi^{-}_{5,x-y}\rangle$ along $(x- y)$.

From this we read off that for a state $|\psi\rangle$ with $M_{x-y}|\psi\rangle=-\ii|\psi\rangle$
the SEV points along $(x-y)$ if it is $d_{z^2}$, $\Gamma_8^{(2)}$ or $\Gamma_7$, and along $(-x+y)$ if it is $d_{x^2-y^2}$ or $\Gamma_8^{(1)}$.
Similarly, for a state $|\psi\rangle$ with $M_{x+y}|\psi\rangle=-\ii|\psi\rangle$
the SEV points along $(x+y)$ if it is $d_{z^2}$, $\Gamma_8^{(2)}$ or $\Gamma_7$, and along $(-x-y)$ if it is $d_{x^2-y^2}$ or $\Gamma_8^{(1)}$.
%
This concludes the set of rules used in the main text.

In addition, the 4-band model requires $s$ and $\Gamma_8^{(2)}$ states: $s$ states behave as $d_{z^2}$ states,
so if $M_x|\psi\rangle=-\ii |\psi\rangle$, the spin points along $+ x$,
if $M_y|\psi\rangle=-\ii |\psi\rangle$, it points along $+ y$,
and if $M_{x-y}|\psi\rangle=-\ii |\psi\rangle$, the spin points along $(+ x- y)$.
As for $\Gamma_8^{(2)}$,
if $M_x|\psi\rangle=-\ii |\psi\rangle$, the SEV points along $+ x$,
if $M_y|\psi\rangle=-\ii |\psi\rangle$, it points along $+ y$,
and if $M_{x-y}|\psi\rangle=-\ii |\psi\rangle$, it points along $(+ x- y)$.
We note that the transformation $x\rightarrow -x$, $y\rightarrow -y$, needed to obtain the hybridization term Eq.~\eqref{v_sf} with all plus signs,
changes the explicit form of $\Gamma_8^{(2)}$ states, but neither their symmetries properties nor their SEV.

\section{Symmetry of surface states at $\bar\Gamma$}

In this section we show how $v$ determines the symmetry of the states which build the $\bar\Gamma$ cone for small momenta.

As usual \cite{liu_ti_model} we introduce  a time-reversal doublet $|\psi_+\rangle$, $|\psi_-\rangle$ which describes surfaces states at $\bar\Gamma$;
it can be written as:
\bea
|\psi_+\rangle&=&\phi_{1}|d_{x^2-y^2}\uparrow\rangle+\phi_{2}|d_{z^2}\uparrow\rangle+\nonumber\\
&&+\phi_{3}|\Gamma_8^{(1)}+\rangle+\phi_{4}|\Gamma_8^{(2)}+\rangle+\phi_{5}|\Gamma_7+\rangle,\label{psi_p}\\
|\psi_-\rangle&=&\phi_{1}|d_{x^2-y^2}\downarrow\rangle+\phi_{2}|d_{z^2}\downarrow\rangle-\nonumber\\
&&-\phi_{3}|\Gamma_8^{(1)}-\rangle-\phi_{4}|\Gamma_8^{(2)}-\rangle-\phi_{5}|\Gamma_7-\rangle.\label{psi_m}\\
\eea
It is constructed such that
\be
\langle(\sigma_x, \sigma_y,  \sigma_z)\rangle =   (a \hat s_x, a  \hat s_y, b \hat s_z),
\ee
with $a$ and $b$ real numbers, as can be seen through the use of Table \ref{table_spin}, and
where we suppose that the $\Gamma_8$ coefficients $\phi_{3}$, $\phi_{4}$ and the $\Gamma_7$ one $\phi_{5}$ are not non-zero at the same time.
The eigenstate with positive energy for the Dirac Hamiltonian $H_{\bar\Gamma}=v_0w (k_x\sigma_y- k_y\sigma_x)$ with $w=+1$ is now
\bea
|\phi^+(\theta_\bk)\rangle=\frac{1}{\sqrt{2}}(|\psi_+\rangle+\ii\e^{\ii\theta_\bk}|\psi_-\rangle),\\
\cos\theta_\bk=\frac{k_x}{k}, \hspace{10pt} \sin\theta_\bk=\frac{k_y}{k}.
\eea
This state has the properties
\bea
M_{y}|\phi^+(0)\rangle&=&-\ii|\phi^+(0)\rangle,\\
M_{x}|\phi^+(\pi/2)\rangle&=&+\ii|\phi^+(\pi/2)\rangle,\\
M_{x-y}|\phi^+(\pi/4)\rangle&=&-\ii|\phi^+(\pi/4)\rangle, \hspace{5pt} \phi_2,\phi_4=0,\\
M_{x-y}|\phi^+(\pi/4)\rangle&=&+\ii|\phi^+(\pi/4)\rangle, \hspace{5pt} \phi_1,\phi_3,\phi_5=0.~~~
\eea
We now see that when $w=+1$, $v=+1$, we must have $M_{x-y}|\phi^+(\pi/4)\rangle=+\ii|\phi^+(\pi/4)\rangle$,
so, in this case, the $\bar\Gamma$ cone is entirely composed by $d_{z^2}$ and $\Gamma_8^{(2)}$;
however, when $w=+1$, $v=-1$, we must have $M_{x-y}|\phi^+(\pi/4)\rangle=-\ii|\phi^+(\pi/4)\rangle$,
so the $\bar\Gamma$ cone is now entirely composed by $d_{x^2-y^2}$, $\Gamma_8^{(1)}$ and $\Gamma_7$.
For $w=-1$ all mirror eigenvalue signs and spin directions are reversed, but the results are the same.
We conclude that $v$ dictates the symmetry of the surface states at $\bar\Gamma$ for small momenta.

\bibliographystyle{apsrev4-1}
\bibliography{tki}